\pdfoutput=1

\documentclass[journal]{IEEEtran}
\ifCLASSOPTIONcompsoc
  \usepackage[nocompress]{cite}
\else
  \usepackage{cite}
\fi
\ifCLASSINFOpdf
  \usepackage[pdftex]{graphicx}
\else
\fi

\newcommand{\etal}{{\em et al\,.}}       
\newcommand{\eg}{{e.g.}}           
\newcommand{\ie}{{i.e.}}           

\begin{document}
%
\title{HF-UNet: Learning Hierarchically Inter-Task Relevance in Multi-Task U-Net for Accurate Prostate Segmentation in CT images}
%
%
%
%

\author{Kelei He,~Chunfeng Lian,~Bing Zhang,~Xin Zhang,~Xiaohuan Cao,~Dong Nie,~Yang Gao,\\Junfeng Zhang,~Dinggang Shen,~\IEEEmembership{Fellow,~IEEE}
\IEEEcompsocitemizethanks{\IEEEcompsocthanksitem K. He and J. Zhang are with Medical School of Nanjing University, Nanjing, P. R. China. Y. Gao are with the State Key Laboratory for Novel Software Technology, Nanjing University, P. R. China. K. He, Y. Gao and J. Zhang are also with the National Institute of Healthcare Data Science at Nanjing University, P. R. China. B. Zhang, X. Zhang are with Department of Radiology, Nanjing Drum Tower Hospital, Nanjing University Medical School, P. R. China. X. Cao is with Shanghai United Imaging Intelligence Co., Ltd. 
\IEEEcompsocthanksitem C. Lian, D. Nie and D. Shen are with Biomedical Research Imaging Center, University of North Carolina, Chapel Hill, NC, U.S.. D. Shen is also with Department of Brain and Cognitive Engineering, Korea University, Seoul 02841, Republic of Korea.}
\thanks{* Corresponding authors: Junfeng Zhang (jfzhang@nju.edu.cn); Dinggang Shen (dgshen@med.unc.edu)}
}

%
%

\markboth{Journal of \LaTeX\ Class Files,~Vol.~14, No.~8, August~2015}%
{Shell \MakeLowercase{\textit{et al.}}: Bare Demo of IEEEtran.cls for Computer Society Journals}
%



\IEEEtitleabstractindextext{%
\begin{abstract}
Accurate segmentation of the prostate is a key step in external beam radiation therapy treatments. In this paper, we tackle the challenging task of prostate segmentation in CT images by a two-stage network with 1) the first stage to fast localize, and 2) the second stage to accurately segment the prostate. To precisely segment the prostate in the second stage, we formulate prostate segmentation into a multi-task learning framework, which includes a main task to segment the prostate, and an auxiliary task to delineate the prostate boundary. Here, the second task is applied to provide additional guidance of unclear prostate boundary in CT images. Besides, the conventional multi-task deep networks typically share most of the parameters (\ie, feature representations) across all tasks, which may limit their data fitting ability, as the specificities of different tasks are inevitably ignored. By contrast, we solve them by a hierarchically-fused U-Net structure, namely HF-UNet. The HF-UNet has two complementary branches for two tasks, with the novel proposed attention-based task consistency learning block to communicate at each level between the two decoding branches. Therefore, HF-UNet endows the ability to learn hierarchically the shared representations for different tasks, and preserve the specificities of learned representations for different tasks simultaneously. We did extensive evaluations of the proposed method on a large planning CT image dataset, including images acquired from 339 patients. The experimental results show HF-UNet outperforms the conventional multi-task network architectures and the state-of-the-art methods. 
\end{abstract}

\begin{IEEEkeywords}
Multi-Task Learning, Segmentation, Prostate Cancer, Boundary-Aware, Attention
\end{IEEEkeywords}}

\maketitle

\IEEEdisplaynontitleabstractindextext

%
\IEEEpeerreviewmaketitle


%
%
%
%


\section{Introduction}
\label{sec:introduction}
\IEEEPARstart{E}{xternal} beam radiation therapy (EBRT) is one of the most commonly used treatments for prostate cancer, the second most common in American men \cite{ps}. Accurate segmentation of the prostate is a very important step in the EBRT planning stage, to maximize the delivery of radiation dose in tumor tissues while avoiding damages to the surrounding healthy organs. Considering manual delineation of the prostate is challenging and time-consuming even for experienced radiation oncologists, developing automated methods (\eg, machine learning-based approaches \cite{guo2016deformable,gao2016accurate,shao2015locally}) to this end is thus of great clinical value. 

\begin{figure}[!t]
  \centering
  \includegraphics[width=\linewidth]{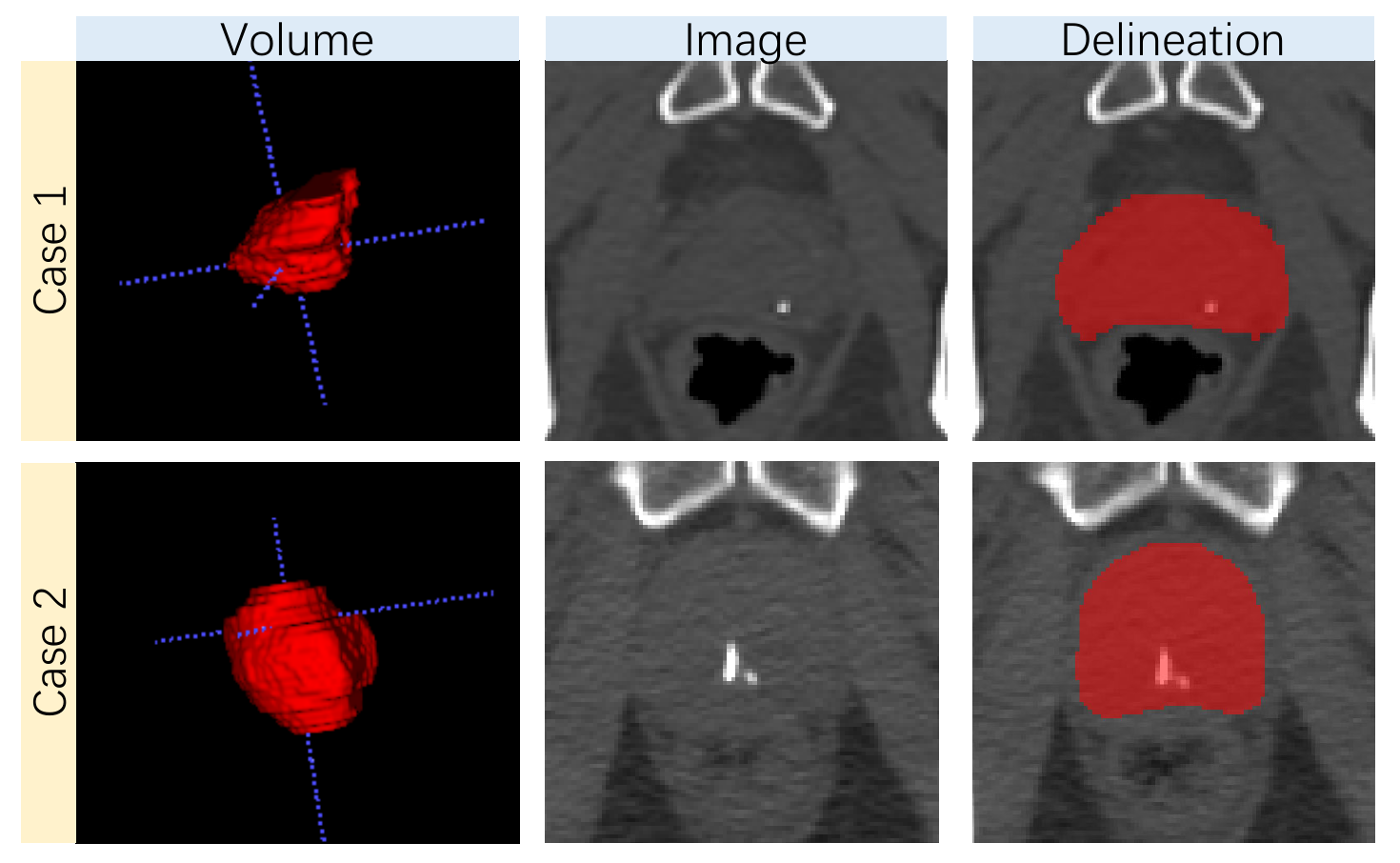}
  \caption{\label{fig:procomp} 3D rendering of segmented prostate volume (left column), cross-sectional image (middle column) and the corresponding delineation (right column) of two sample patients. The examples show low soft tissue contrast (see middle column), and large inter-subject variance of prostate shapes (see right column) in the CT images.}
\end{figure}

To date, fully convolutional network (FCN) \cite{long2015fully}, as well as its variants \cite{ronneberger2015u,nie2016fully,milletari2016v,chen2018voxresnet}, have been successfully applied for medical image segmentation. 
These methods show the state-of-the-art performance due to task-oriented extraction and integration of both semantic knowledge and local details for pixel-wise dense predictions. 
However, the direct application of the conventional FCNs for prostate segmentation is challenging, mainly because 1) the prostate boundaries cannot be easily distinguished in CT images due to the low tissue contrast, and 2) the organ shapes and appearances in CT images usually show large variance across different subjects. (see Fig. \ref{fig:procomp}).

\begin{figure*}[!t]
  \centering
  \includegraphics[width=\linewidth]{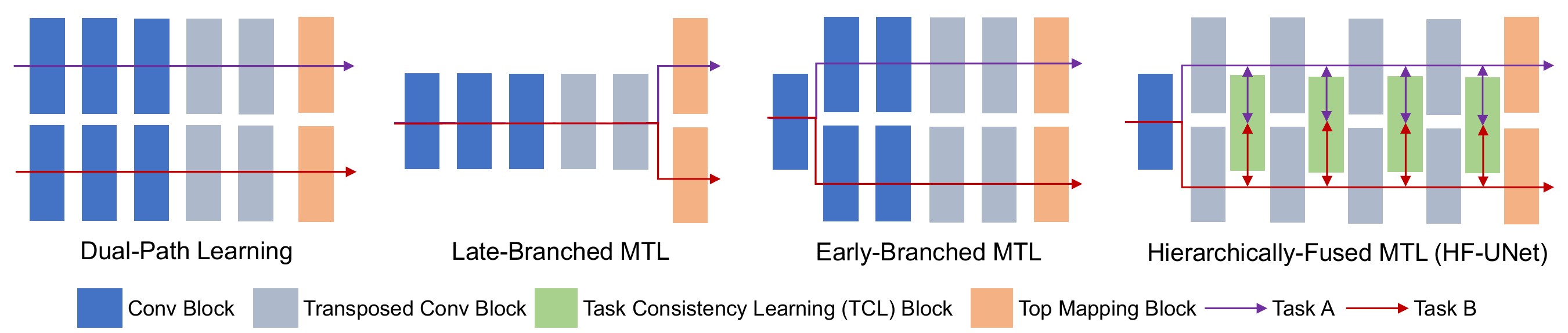}
  \caption{\label{fig:modelcomp} Comparison of 1) dual-path learning, 2) early-branched multi-task learning (MTL), 3) late-branched MTL, and 4) our proposed hierarchically fused MTL U-Net (HF-UNet). The color rectangles denote different kinds of blocks, and the arrows indicate the data flow for the two tasks. Typically, the dual-path learning tackle the two tasks with separate models; while for MTL, the late-branched network branches are built to two sub-tasks in the top-mapping layer, while the early-branched network branches are built after initial convolutional block.}
\end{figure*}

Also, due to the down-sampling operations used in conventional FCNs, the spatial correlations between voxels located at the prostate boundary are usually destroyed, and thus cause the final segmentations failing to segment complete prostate contours. 
To tackle these challenges, for learning-based methods, a potentially feasible solution is to leverage the context-aware strategy \cite{gao2016accurate,shao2015locally,chen2017dcan,he2019pelvic} in the multi-task learning framework to automatically refine the organ boundaries during segmentation. For example, Shao \etal \cite{shao2015locally} proposed a landmark-based contour detection method for segmenting the prostate and rectum with a deformable model. Notably, for FCN, Chen \etal \cite{chen2017dcan} proposed a contour-aware network for gland segmentation, where the delineation of gland contours (\ie, contour vs. non-contour voxel classification) is incorporated as a supplementary task in addition to the main task of gland segmentation. However, the disadvantage of such operation is that, as the prostate boundaries in CT images are unclear, regarding their delineation as a binary classification task is practically unreliable.
Moreover, it is worth mentioning that most existing deep multi-task networks (\eg, \cite{chen2017dcan,zhang2014facial}) are implemented in a very similar way. That is, they usually consist of 1) a backbone with shared weights to learn common features for all tasks, which is further followed by 2) task-specific layers that are branched for different tasks. Then, based on the choice of branched levels, similar to fusion strategies of multi-modality networks \cite{snoek2005early,feichtenhofer2016convolutional}, conventional multi-task networks also can be concluded to two formulations, i.e., the early-branched network and the late-branched network. (See the comparison of different multi-task models in Fig. \ref{fig:modelcomp}.) To make extra use of multiple guidance, the late-branched network is a common choice in recent studies. For example, Liu \etal \cite{liu2017deep} proposed to predict brain disease and regress clinical scores simultaneously via a multi-task multi-channel network. 
Intuitively, since nearly all network parameters are shared across all tasks, the late-branched network aggressively assumes that different tasks could be commonly handled by the same feature representation. In open set applications, such strong constraint is inflexible, as it inevitably ignores the specificities of different tasks.

To overcome the strong constraint, we propose a novel multi-task fully convolutional network, namely HF-UNet, for automatic segmentation of the prostate in CT images. Specifically, we regard both the manually-delineated prostate boundaries and adjacent voxels as valuable task-related knowledge for localization of unclear prostate boundary in CT images. Then, instead of formulating organ boundary delineation as a binary classification problem \cite{chen2017dcan}, we integrate the delineation of the morphological representation of the prostate boundary as an auxiliary task for prostate segmentation (\ie, the main task). Accordingly, the proposed HF-UNet is composed of two branches to infer the predictions of the aforementioned two tasks with an encoder-decoder structure. As shown in Fig. \ref{fig:oass}, each branch of our network is U-Net like, for which is further composed of three cascaded down-sampling blocks and three cascaded up-sampling blocks hierarchically defined at different levels, with each block consisting of several convolutional layers and a size-transform operation (\ie, the transposed convolutional layer). Thus, the same-level blocks in the two different branches learn with the same receptive field of an input image. To encourage feature representations learned by the two branches to be complementary to each other, while not forcing them to be identical, the same-level blocks in the two different branches are communicated via a novel dual attention-based Task Consistency Learning (TCL) module. Then, by inserting these TCL blocks at multiple levels, the two complementary branches can share task-oriented information hierarchically. Obviously, we here use a looser architecture that obeys basic transfer learning rules in this case, by making a trade-off between the early-branched and the late-branched multi-task networks when dealing with two tasks. (See in Fig. \ref{fig:modelcomp}). 
Thus, the flexibility of the model can be enhanced. Moreover, the main idea of minimizing the discrepancy of two distributions (\ie, the features for segmentation and contour delineation task) has been well demonstrated in the related research area of domain adaptation \cite{ben2007analysis,ganin2016domain}, which aims to minimize the discrepancy of features for two input domains.

The contributions of this paper are three-fold. 
(1) We propose a hierarchically fused multi-task architecture, which can better hierarchically learn the complementary features for the two tasks. Specifically, the proposed TCL blocks are proposed to implement hierarchical feature interaction for solving two tasks. 
(2) We tackle prostate segmentation task by incorporating contour awareness into the network, and solve it via multi-task learning. The network is guided by regressing a specially designed contour sensitive label concerning the prostate boundary and their contexts. 
(3) We evaluate our proposed method on a large 339-patient CT image dataset, showing the improved performance of prostate segmentation compared with state-of-the-art methods.

\section{Related Work}
In the following, we review related literature in two aspects: 1) the application of \emph{prostate segmentation in CT images}; 2) the method of \emph{multi-task deep networks}.

\subsection{Prostate Segmentation in CT Images}
Several methodologies are developed for prostate segmentation \cite{guo2016deformable,gao2016accurate,shao2015locally}. Among these methods, deformable models are a series of computer algorithms that model the variability of a certain class of objects \cite{mcinerney1996deformable}. Since these methods leverage the statistics to deform a shape representation on a specific class of objects, they can well describe organ contours in medical image conditions, especially in CT images. Thus, the most advanced segmentation methods for prostate in CT images are often based on deformable models. For example, Gao \etal \cite{gao2016accurate} proposed to use a deformable model to regress organ displacement maps. Shao \etal \cite{shao2015locally} proposed a landmark-based contour detection method for segmenting the prostate and rectum with a deformable model. Notably, these advanced methods often design strategies specifically focused on distinguishing the unclear prostate boundary, and solving this task by leveraging the power of multi-task learning. This shows the importance of these two methodologies. However, these methods were mostly developed on hand-crafted features. Recently, deep neural networks have been adopted in various medical image analysis applications \cite{ronneberger2015u,nie2016fully}, because of their task-oriented learning ability. Our proposed network structure also belongs to the group of FCNs to address the challenging segmentation task in CT images. In general, we leverage both the power of deep learning and multi-task learning in this work. Specifically, the delineation of the contour is adopted as an auxiliary task.

\subsection{Deep Multi-Task Neural Networks}
The effectiveness of incorporating multi-task learning with the deep neural network has been demonstrated in various computer vision applications, \ie, facial landmark detection \cite{zhang2014facial}, speech synthesis \cite{wu2015deep}, and human pose estimation \cite{li2014heterogeneous}. Also, several studies have adopted this learning paradigm in medical image analysis. For example, Chen \etal \cite{chen2017dcan} proposed a multi-task deep network for gland segmentation with contour awareness. Liu \etal \cite{liu2017deep} proposed to predict brain disease and regress clinical scores simultaneously via a multi-task multi-channel network.
In general, their network structures can be formulated into the abovementioned fashion, which uses one base network for feature sharing through all target domains. However, the efficacy of this methodology is limited as no task-specific feature is preserved. To solve this problem, we loosen this constraint to explore a more flexible network architecture. Its core assumption is to enable self-domain information being preserved by designing single paths for different tasks, and to achieve learned feature sharing by inserting inter-network modules.

\begin{figure*}[!t]
  \centering
  \includegraphics[width=\textwidth]{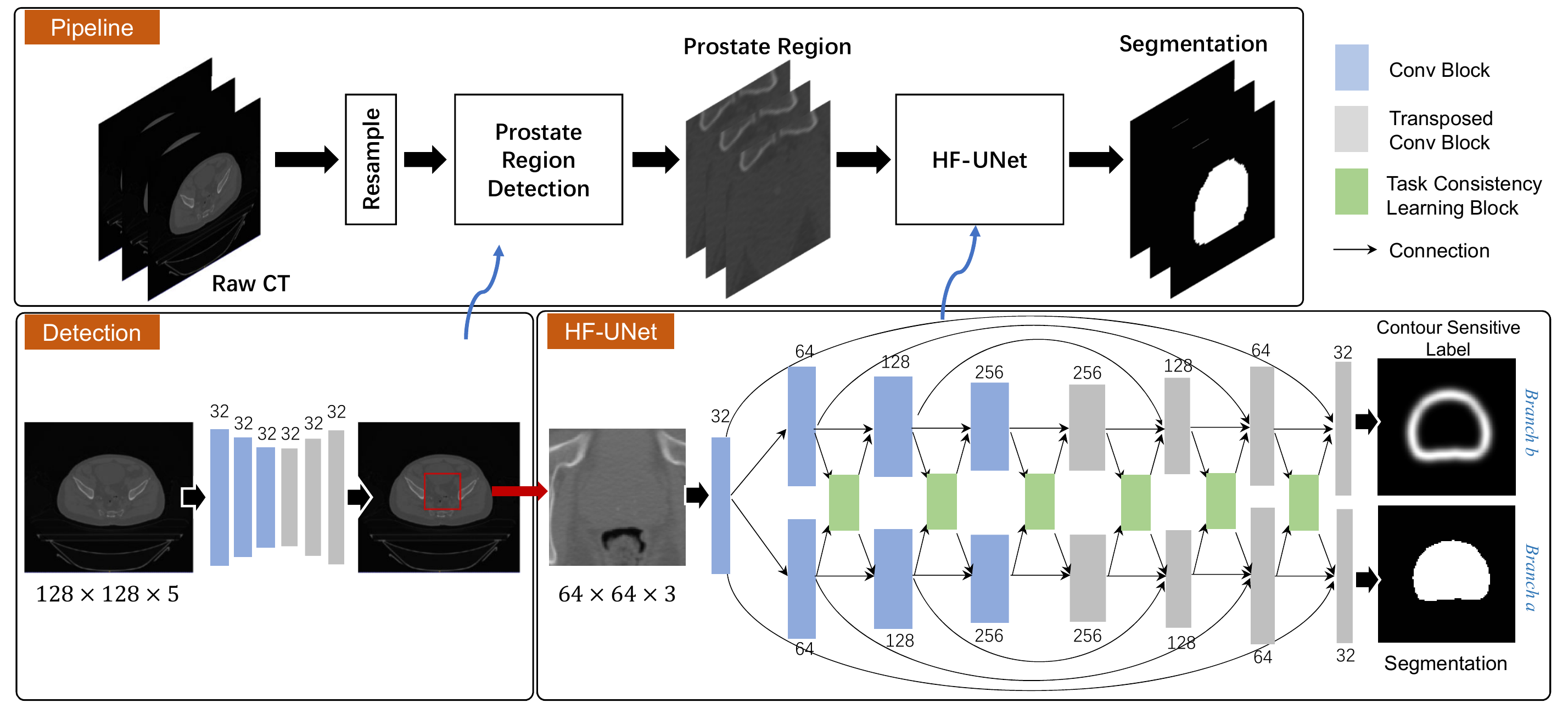}
  \caption{The pipeline of the proposed segmentation framework. The architecture of HF-UNet-6 contains six TCL blocks as shown in the lower-right box. The network has two branches for solving two tasks, i.e., 1) contour sensitive label regression and 2) prostate segmentation. The TCL blocks (denoted as green boxes) are used to pass interactive information between the two branches. The two tasks can be solved end-to-end by the proposed architecture.}
  \label{fig:oass}
\end{figure*}

\section{Methods}

The prostate is located in a relatively small region, compared with the whole pelvic CT image. Thus, large background noise will weaken the network. To address this issue, we propose a two-stage segmentation framework, with its pipeline shown in Fig. \ref{fig:oass}.


\subsection{Automatic Prostate Region Detection}
We first adopt a region localization network to crop a sub-image that includes the whole prostate region completely, as shown in Fig. \ref{fig:oass}. 
Distinct to semi-automatic methods, \eg, \cite{shi2015semi}, our proposed framework can automatically localize prostate region by a network with a U-Net \cite{ronneberger2015u} architecture. Afterwards, the center of the prostate is determined based on the coarsely predicted segmentation map. We crop a size of $128 \times 128 \times 128$ region based on the obtained organ center to fully cover the prostate organ. Then, the patches cropped from the identified region are used to train the subsequent network (i.e., HF-UNet), instead of the raw CT images, to perform more accurate segmentation of prostate and more efficient of network training.



\subsection{HF-UNet}
The architecture of our proposed HF-UNet is shown in Fig. \ref{fig:oass}. Basically, we set two tasks to HF-UNet for accurate prostate segmentation: 1) segmentation of the prostate (denote as $Branch_a$), and 2) delineation of the prostate boundary (denote as $Branch_b$). Here, the second task (\ie, the task for contour-aware sub-network) is introduced to provide critical morphological guidance for the network, which can help well distinguish the prostate boundary. 
Due to the down-sampling operations used in conventional FCNs, the spatial correlations between voxels located at the prostate boundary are usually destroyed, which causes the final segmentations failing to delineate complete prostate contours. 
To address this issue, the idea of incorporating organ contours as complementary information for learning-based methods has been presented and demonstrated in several studies \cite{shao2015locally,gao2016accurate,chen2017dcan}. 
For example, Chen \etal \cite{chen2017dcan} proposed a deep multi-task network with the guidance of gland contours to improve segmentation performance. 
Previous methods \cite{chen2017dcan,shen2017boundary} often formulate the delineation of the organ boundary as a binary classification task, by simply treating the voxels of the organ contours as positive samples, and others as negative samples. However, the prostate boundaries are hard to be distinguished, making these methods less applicable. The voxels near the prostate boundaries contain abundant contextual information for the contour pixels. Moreover, manually delineated prostate contours are often not reliable. 
Previous methods \cite{chen2017dcan} often formulate the delineation of the organ boundary as a binary classification task, which is not reasonable.
Therefore, we assume that the voxels near the organ boundary are also task-relevant. 
\subsubsection{Delineation of Organ Boundaries} 
In this work, we solve the delineation task as a regression problem with the consideration of the pixels near the contour as sub-groundtruths. Inspired by the idea presented in \cite{Pfister_2015_ICCV}, we formulate each contour point and its surrounding pixels into a Gaussian distribution, with a kernel of $\sigma$ (\ie, $\sigma=5$ here). 
Formally, we denote the input of the loss layer as $(\xi_X,y)$, where $\xi_X$ is the output feature of the network and $y$ denotes the ground-truth label of the prostate contour. We use mean squared error to calculate the difference between the two tasks. Then the objective of contour-aware w.r.t. the inputs $(X,y)$ and weight $\theta$ can be written as,

\begin{equation}{}
\begin{array}{rrclcl}
\displaystyle \mathcal{L}_{reg.} = \mathop{\arg\min}_{\theta} \sum_{(\xi_X,y)}{\sum_{i,j}{||S(y_{ij})-\phi(\xi_X,\theta)||^2}},
\end{array}
\label{eq:Reg}
\end{equation}

where $\phi(X,\theta)$ is the output of the contour-aware branch.
And $S(y_{ij})$ is the aggregation of the Gaussian filters, where $y$ stands for the position $(i,j)$, which can be formulated as,

\begin{equation}{}
\begin{array}{rrclcl}
\displaystyle S(y_{ij})=\sum_{k \in K,l \in L}{\frac{1}{\sigma \sqrt{2\pi}}e^{\frac{(i-k)^2+(j-l)^2}{2\sigma^2}}},
\end{array}
\label{eq:sigma}
\end{equation}

where $(K,L)$ denotes the neighbor set of $(i,j)$ with the distance smaller than $\sigma$.


\begin{figure}[!h]
  \centering
  \includegraphics[width=\linewidth]{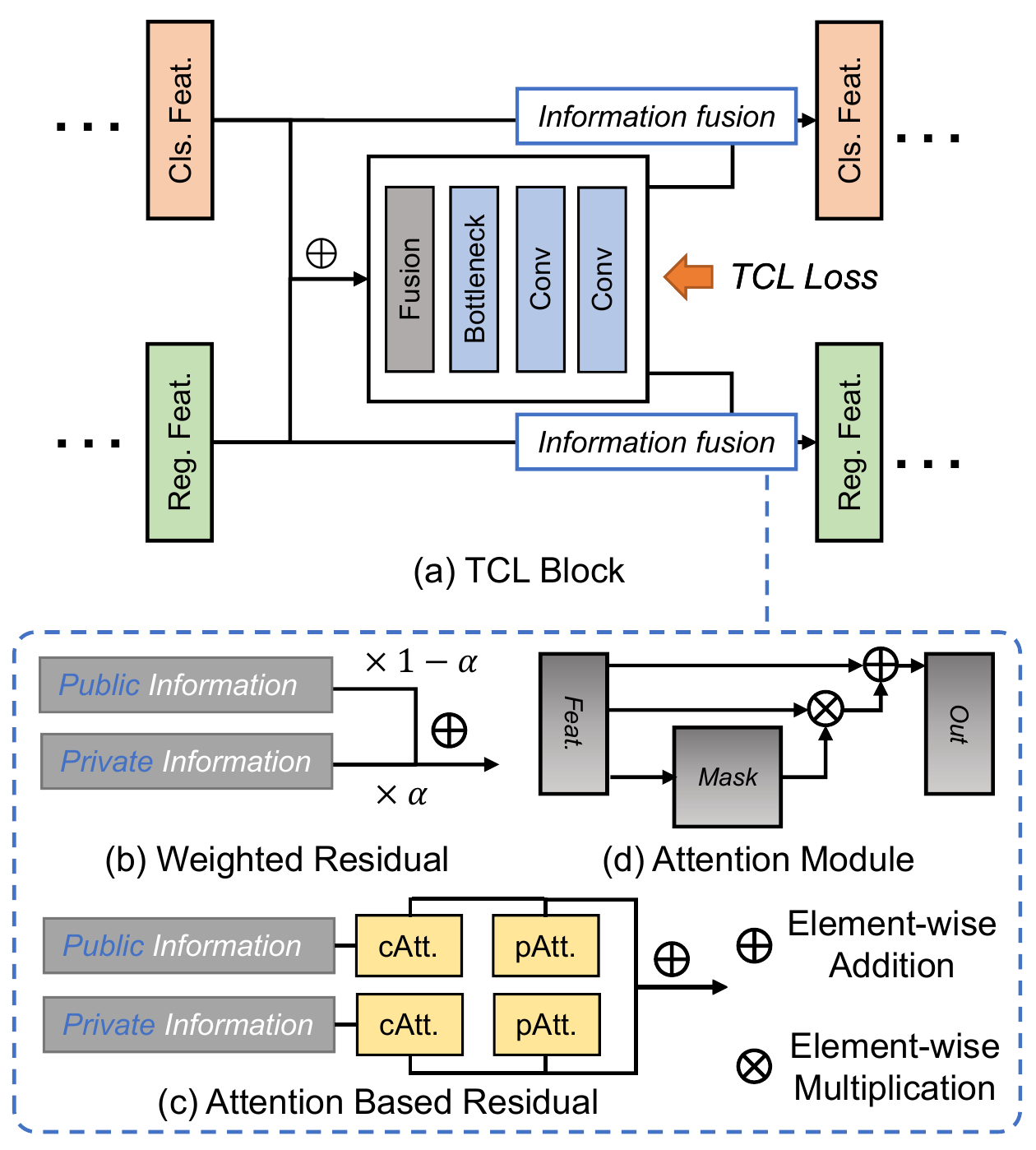}
  \caption{\label{fig:aTCLblock}Illustration of the weighted residual Task Consistency Learning (TCL) block and the attention-based TCL block. 'Cls. Feat.' denotes the intermediate features in the classification branch, and 'Reg. Feat.' denotes the intermediate features in the regression branch. In weighted residual TCL block, the shared information and task-specific information are balanced by weight ratio $\alpha$, and, in the attention-based TCL block, the two kinds of information are activated by both the channel-wise attention (denoted as cAtt.) and position-wise attention (denoted as pAtt.) modules.}
\end{figure}

\subsubsection{Task Consistency Learning (TCL) Block}
The information of two branches is combined and learned by the proposed TCL blocks. Herein, the TCL block can be regarded as modeling the knowledge learned from both branches from a global perspective (\ie, task-level). And the features learned in each branch can be regarded as the knowledge learned from a local perspective (\ie, sample-level). 

We build two kinds of TCL blocks: 1) the weighted residual TCL block to evaluate the effectiveness of the shared information and task-specific information, and 2) the attention-based TCL to adaptively learn the ratio of the aforementioned two kinds of information, as shown in Fig. \ref{fig:aTCLblock}. 
For the convenience of computation, we use the summation operation to combine information from different sources, inspired by the residual learning setting in \cite{he2016deep}. Then, it is followed by several convolutional layers for information extraction. In fact, other operations such as concatenation can also be applied. 
The fusion feature will be guided by a task consistency loss to merge the gap between the two different tasks. In this work, we use the aggregation of mean squared error (MSE) of the feature pairs between TCL blocks and the two branches. Letting the network to have a total of $M$ levels of TCL blocks, and the features outputted by the segmentation branch, contour-aware branch and TCL block in the $m$th level as $\xi_{seg.,m}$, $\xi_{cont.,m}$, $\xi_{tcl,m}$, the final loss for TCL is the aggregation of losses calculated on the pairs of the two tasks with the fusion feature outputed by TCL block. The loss is then written as follows,

\begin{equation}
\begin{array}{rrclcl}
\displaystyle \mathcal{L}_{tcl.} = \sum_{m=1}^{M}{(\|\xi_{seg.,m}-\xi_{tcl,m}\|_2+\|\xi_{cont.,m}-\xi_{tcl,m}\|_2)},
\end{array}
\label{eq:Tcl}
\end{equation}

The mean squared error can build the consistency of the fusion features with the features from the segmentation and contour delineation sub-networks. 

In common practice, the noise information will be increased after the residual connection. This will be harmful to the training of the network. Following the setting in \cite{Szegedy_2015_CVPR}, 
which has been successfully demonstrated in \cite{he2016deep,huang2017densely}, 
the size of $1\times1$ convolutional layer with the stride of $1$, namely bottleneck layer, can better preserve identical information and thus deduce redundant feature maps. In this paper, we adopt this idea and set each fusion layer followed by a bottleneck layer. After several convolutional layers, the features generated by the TCL blocks will be fed back into each branch. 

To investigate the effectiveness of two kinds of information, we build HF-UNets with weighted residual TCL blocks.
The weighted residual connection is used to balance the effectiveness of the information learned in each branch with the TCL blocks. 
Formally, for a certain block in each branch, we define two connections: 1) \emph{private connection}, \ie, the connections between the sub-network blocks; 2) \emph{public connection}, \ie, the connection between the sub-network and the TCL block. Then, the public information in the $m$th level (denoted as $\xi_{public,m}$) and the private information in the $m$th level (denoted as $\xi_{private,m}$) is defined accordingly. 
Public information and private information are aggregated by residual connections. 
Moreover, to investigate the influences of public information and private information, we introduce a weight $\alpha$ for constructing the weighted residual operation. (See Fig. \ref{fig:aTCLblock}) Then, the input feature of the $m+1$th level in each branch (denoted as $\xi_{private,m+1}^{in}$) can be written as,

\begin{equation}
\begin{array}{rrclcl}
\displaystyle \xi_{private,m+1}^{in} = \alpha\xi_{private,m} + (1-\alpha) \xi_{public,m}, 
\end{array}
\label{eq:Ratio}
\end{equation}

Obviously, the smaller $\alpha$ indicates that more public information is fed into each branch, and vice versa. When $\alpha$ is as small as zero, the network will degenerate into a fully-shared network (i.e., the late-branched MTL network in Fig. \ref{fig:modelcomp}), and will change to dual-path learning network (in Fig. \ref{fig:modelcomp}) if $\alpha=1.0$. The channels of the inputs are often roughly concatenated in the existing works, \ie, using $\alpha=0.5$. In this work, we regard $\alpha$ as a hyper-parameter and explore the importance of self-learned information and shared information by adjusting this ratio $\alpha$. The experiments show the performance is affected by $\alpha$.
This inspires us to design the TCL blocks with an automatically learning strategy. We regard the ratio of the information being controlled by both the channel-wise and position-wise of the features. Thus, the channel-wise attention module (cAtt.) and the position-wise attention module (pAtt.) are performed to balance the weight of the two kinds of information. We connect each kind of information of the two attention modules by residual connections. Let the feature $\xi$ have the dimension of $H \times W \times d$, $\{A_{c1},A_{c2}\}$ denote the $d\times d$ dimensional attention mask of the cAtt. module, and $\{A_{p1},A_{p2}\}$ denote the $H\times W \times H \times W$ dimensional attention mask of the pAtt. module. Therefore, the output feature by weighted residual connection can be written as,

\begin{equation}
\begin{array}{rrclcl}
\displaystyle \xi_{private,m+1}^{in} = A_{c1}\times\xi_{private,m} + \xi_{private,m} \\
+ A_{c2}\times\xi_{public,m} + \xi_{public,m} \\
+ A_{p1}\times\xi_{private,m} + \xi_{private,m} \\
+ A_{p2}\times\xi_{public,m} + \xi_{public,m}.
\end{array}
\label{eq:Over}
\end{equation}


\subsubsection{Multi-Task Learning} 
We solve the aforementioned three tasks (i.e., segmentation, regression and task consistency learning) jointly via multi-task learning. The three tasks can provide different kinds of information cues to guide the training of HF-UNet. The final objective function can be written as follows,

\begin{equation}
\begin{array}{rrclcl}
\displaystyle \mathcal{L} = \lambda_1 \mathcal{L}_{cls.} + \lambda_2 \mathcal{L}_{reg.} + \lambda_3 \mathcal{L}_{tcl} + R(\Theta),
\end{array}
\label{eq:Obj}
\end{equation}

where $L_{cls.}$ denotes the loss of classification task for segmenting the prostate, and $L_{reg.}$ denotes the loss of the regression task for delineating the prostate contours. The weights $\Theta = \{\theta_{cls.},\theta_{reg.},\theta_{tcl}\}$ are regularized by $R(\cdot)$ as typically used in the common deep neural networks. We use cross-entropy loss for segmentation task. Letting ${\hat{p}}$ denote the predicted probability, and $y$ denote the ground-truth label, the loss $\mathcal{L}_{cls.}$ can be thus written as,

\begin{equation}
\begin{array}{rrclcl}
\displaystyle \mathcal{L}_{cls.} = -(y\log{\hat{p}}+(1-y)\log{1-\hat{p}}),
\end{array}
\label{eq:Cls}
\end{equation}

The loss in Eq. (\ref{eq:Reg}) is used for regression. The regression task often generates a large magnitude of losses. Therefore, we employ two tricks to address this issue. First, we only train the parameters of the segmentation branch in the cold start phase, by fixing all the parameters in the regression branch and TCL blocks. This can better initialize the weights for unbalanced multi-task learning. Second, we set unbalanced weights for three losses. In this work, we set the weight factor $\lambda_1 = 1.0 $, $\lambda_2 = 0.01$ and $\lambda_3 = 1.0$.

\section{Experimental Results}
\subsection{Dataset} We evaluate the proposed method on a large planning CT image dataset consisting of 339 patients. The dataset contains 339 images, and the image size is $512\times512\times(61\sim508)$. These images have the in-plane resolution of $0.932 \sim 1.365$ mm, and the slice thickness of $1\sim3$ mm.
This dataset is challenging, due to the following three reasons: 1) The CT images have very low tissue contrast, making the prostate very hard to distinguish. 2) The appearance and shape vary greatly across subjects. 3) Only one planning image avaliable for one patient.

\subsection{Implementation Details} Our method was implemented on the popular opensource framework \emph{PyTorch} \cite{paszke2017automatic}. The experiments are performed on a workstation with an NVidia 1080Ti GPU.
We randomly separated the dataset into training ($70\%$), validation ($10\%$) and testing ($20\%$) subsets.
For image preprocessing, we first resample all images to have the same resolution $\{1:1:1\}\mathrm{mm}^{3}$, since the images are captured by different CT scanners from different manufacturers.
For each image, we preserve part of the CT image that only contains the patient's body through a threshold-based cropping method, in order to reduce the noisy background. 
We further perform intensity normalization for all images, and the intensity values are normalized into $[-1,1]$.
We generate input images for the proposed network by cropping 2-D image slides using a sliding window strategy. In the \emph{training} stage, we randomly crop patches in the region of the prostate. In the \emph{testing} stage, we apply the trained convolutional filters directly to the region image (i.e., the $128\times128\times128$ image), to get the full segmentation of the whole region. The patch size in our work is $64\times64\times3$, while the batch size for training is set to $64$. Here, we input sequential slices to predict the label of the \emph{middle} slice. Such a strategy incorporates the 3-D spatial relationship of the under-predicted slice, thus helping enhance the prediction performance.
The convolutional layers in these networks are followed by a rectified linear unit (ReLU) as the activation function. The networks in experiments are optimized by Stochastic Gradient Descent (SGD) algorithm. We train the networks using $60$ epochs. The learning rate is decreased from $0.01$ to $0.0001$ by a step size of $2\times10^5$. 

\subsection{Metrics}

We utilize four commonly used metrics, i.e., Dice similarity coefficient (DSC), average surface distance (ASD), sensitivity (SEN) and positive predictive value (PPV) to evaluate the performance of our proposed method. These four metrics are defined as follows,

1) DSC:

\begin{equation}
\begin{array}{rrclcl}
\displaystyle DSC=\frac{2\|Vol_{gt}\cap Vol_{seg}\|}{\|Vol_{gt}\|+\|Vol_{seg}\|}
\end{array}
\label{eq:DSC}
\end{equation}

2) ASD:

\begin{equation}
\begin{array}{rrclcl}
\displaystyle ASD=\frac{1}{2}\{\mathrm{mean}\min\limits_{a\in Vol_{gt},b\in Vol_{seg}}{d(a,b)} \\
        + \mathrm{mean}\min\limits_{a\in Vol_{seg},b\in Vol_{gt}}{d(a,b)} \}
\end{array}
\label{eq:ASD}
\end{equation}

3) SEN and PPV:
\begin{equation}
\begin{array}{rrclcl}
\displaystyle PPV=\frac{\|Vol_{gt} \cap Vol_{seg}\|}{\|Vol_{seg}\|};
\end{array}
\hspace{1em}
\begin{array}{rrclcl}
\displaystyle SEN=\frac{\|Vol_{gt} \cap Vol_{seg}\|}{\|Vol_{gt}\|}.
\end{array}
\label{eq:PPV} 
\end{equation}

where $Vol_{gt}$ denotes the voxel set of groundtruth, and $Vol_{seg}$ denotes the voxel set of the segmentation. $d(a,b)$ denotes the Euclidean distance of $\{a,b\}$.

\subsection{Ablation Study}

\subsubsection{Comparison with Different Multi-Task Models}
To evaluate the effectiveness of the proposed contour-aware method, we composed the conventional U-Net (denoted as "U-Net") that only perform the segmentation task, and the two conventional multi-task deep networks, i.e., the late-branched network (denote as "LB") and the early-branched network (denote as "EB"). In the late-branched network, only one conventional U-Net backbone (with the same settings of layers and parameters to U-Net) is kept, followed by two top-mapping blocks containing two convolutional layers (with each block corresponding to a specific task). In the early-branched network, the network is divided into two branches after the initial convolutional block. In each branch, the parameters are set to the same as the conventional U-Net architecture.

To investigate the performance of different configurations of HF-UNets with the different number of TCL blocks, we construct three types of HF-UNet, in which the weighted residual TCL blocks w.r.t. $\alpha=0.2$ are inserted from the top-level to the intermediate level. The different types of networks are constructed with one TCL block (denoted as "HF-UNet-1"), two TCL blocks (denoted as "HF-UNet-2"), and three TCL blocks (denoted as "HF-UNet-3"). The network architecture is displayed in Fig. \ref{fig:oass}.
To further investigate the effectiveness of feature sharing through the whole network, we construct HF-UNet with six TCL blocks (denote as "HF-UNet-6"). To achieve this goal, we split the feature extraction path into two separate feature extraction paths, which are similar to the inference path; then, we insert the TCL blocks into each level among them. 

\begin{table}[!htbp]
\renewcommand{\arraystretch}{1.3}
\centering
\caption{\label{Table:DifArc} Quantitative comparison of different multi-task network architectures (\ie, U-Net, EB, LB network and HF-UNets) on mean DSC and ASD in the 339-patient CT image dataset. Best results are indicated in \textbf{bold}. 'FB' denote the number of fusion blocks, formating in \{the number of shared parameter blocks + the number of TCL blocks\}. The HF-UNets are trained with $\alpha=0.2$.}
\begin{tabular}{cccc}
\hline
\textbf{Methods} & \textbf{FB}  & \textbf{DSC$\pm$std} & \textbf{ASD$\pm$std(mm)} \\
\hline
\textbf{U-Net} & 0+0 &0.837$\pm$0.063 & 3.892$\pm$2.935 \\
\textbf{EB} & 1+0 & 0.864$\pm$0.046 & 2.384$\pm$1.777 \\
\textbf{LB} & 7+0 & 0.863$\pm$0.049 & 2.196$\pm$1.689 \\
\hline
\textbf{HF-UNet-1} & 6+1 & 0.872$\pm$0.032 & 2.013$\pm$1.387 \\
\textbf{HF-UNet-2} & 5+2 & 0.873$\pm$0.029 & 2.071$\pm$1.412 \\
\textbf{HF-UNet-3} & 4+3 & 0.875$\pm$0.029 & 1.711$\pm$1.007 \\
\textbf{HF-UNet-6} & 1+6 & \textbf{0.878$\pm$0.029} & \textbf{1.366$\pm$0.543} \\
\hline
\end{tabular}\\
\end{table}

The performance in terms of DSC and ASD for these seven methods is reported in Table \ref{Table:DifArc}. For a fair comparison, all these networks are trained by the same hyper-parameters. It can be observed from Table \ref{Table:DifArc} that HF-UNet-6 achieves the best performance among HF-UNet-1$\sim$6, indicating that the segmentation performance was improved by increasing the number of TCL blocks. For example, HF-UNet-6 achieves ASD of $1.362$mm, which is significantly better than that of HF-UNet-1 (\ie, $1.653$mm). This validates our assumption that the hierarchically sharable policy can benefit the feature fusion.

The performance on ASD of the late-branched network is slightly better than the early-branched network, which is also partially demonstrated in other existing works (\eg, \cite{Karpathy_2014_CVPR}).
When comparing with the results of U-Net, the segmentation performance is improved by incorporating the contour awareness to the network (as we do in all the multi-task networks, including early- and late-branched network, and HF-UNets). For example, in terms of ASD, the large margin ($>0.5$mm) between HF-UNet-6 and U-Net clearly demonstrates the efficacy of considering the contour awareness in the network. In terms of DSC, the $1\%$ improvement of HF-UNet-1 over late-branched network indicates that the proposed strategy of learning inter-task relevance used in HF-UNet-1 can generate better segmentation results in comparison to the whole-share parameter strategy used in the late-branched network. Also, HF-UNet-6 yields the overall best performance, indicating the effectiveness of fusion at all inference levels. 

\begin{table*}[!htbp]
\renewcommand{\arraystretch}{1.3}
\centering
\caption{\label{Table:EvaPro}Quantitative Comparisons of mean DSC, ASD, SEN and PPV for prostate segmentation on 339 Planning CT Images. 'Num.' denotes the number of cases included in the dataset. The methods in the second row are reported on the same 339 planning CT images.(Bests are in \textbf{bold})}
\setlength{\tabcolsep}{5pt}
\begin{tabular}{c|c|c|cccc}
\hline
\textbf{Methods}  & \textbf{Num.} & \textbf{Method Type} & \textbf{DSC} & \textbf{ASD} & \textbf{SEN} & \textbf{PPV}\\
\hline
\textbf{Martinez} & 116 & Deformable Model & 0.87 &  - \\
\textbf{Shao} & 70 & Deformable Model& 0.88 & 1.86 & - & - \\
\textbf{Gao} & 29 &Deformable Model& 0.86 & 1.85 & - & - \\
\textbf{Gao} & 313 &Deformable Model& 0.87 & 1.77 & 0.88 & 0.85\\
\hline
\textbf{U-Net} & 339 & Deep Network& 0.84 & 3.89 & 0.87 & 0.81\\
\textbf{V-Net} & 339 &Deep Network& 0.85 & 2.27 & 0.88 & 0.84\\
\textbf{He} & 339 &Deep Network& 0.87 & 1.71 & 0.88 & 0.87\\
\textbf{HF-UNet-6 ($\alpha=0.2$)}  & 339 & Deep Network& 0.87(8) & 1.36 & 0.88 & 0.88 \\
\textbf{HF-UNet-6-cAtt}  & 339 & Deep Network& 0.87(3) & 1.34 & 0.85 & 0.89 \\
\textbf{HF-UNet-6-pAtt}  & 339 & Deep Network& 0.87(7) & 1.40 & 0.88 & 0.88 \\
\textbf{HF-UNet-6-dAtt}  & 339 & Deep Network& \textbf{0.88(0)} & \textbf{1.31} & \textbf{0.88} & \textbf{0.89} \\
\hline

\end{tabular}
\end{table*}

\subsubsection{Investigation and Visualization of the Features in HF-UNets}

\begin{figure}[!htbp]
  \centering
  \includegraphics[width=\linewidth]{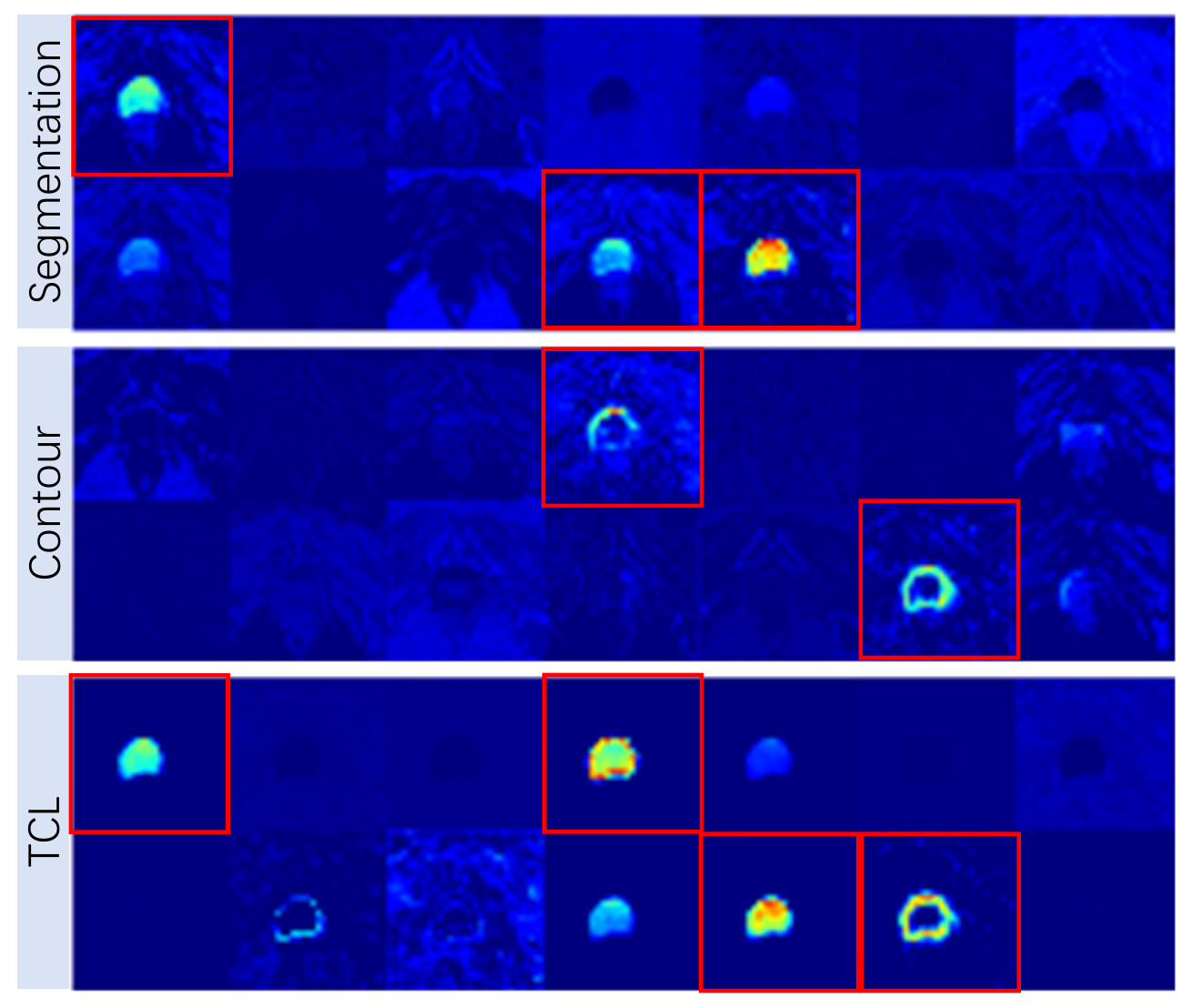}
  \caption{\label{fig:block6} Visualization of the features in the top-level blocks of the segmentation branch (i.e., $branch_a$), the contour-aware branch (i.e., $branch_b$), and the TCL block. Due to limit of space, we convert the first 14 channels of the 3-D features (i.e., $32\times64\times64$) into 2-D map by channels for better visualization. The color from blue to red indicates low to high activity of the neurons.}
\end{figure}

To investigate the learned features at different levels of HF-UNet-6, we visualize outputs of top-level blocks in Fig. \ref{fig:block6}. Different from the late-branched network that shares parameters at almost all levels, HF-UNet has two branches to deal with two specific tasks. Thus, we visualize the features learned from the top-block in the two branches of HF-UNet-6, with the features from the corresponding TCL block. Fig. \ref{fig:block6} shows that high activation of neurons in several channels (marked by red boxes) appears in a large area which can indicate the complete organ segmentation in the segmentation branch, and also appears mainly on organ boundaries in the contour-aware branch. This demonstrates that our HF-UNet method can learn task-oriented features for performing different tasks, which is particularly useful for multi-task learning problems. Moreover, feature maps of TCL block contain both information of the two branches, which demonstrate the effectiveness of the proposed inter-task learning assumption. Besides, the activation in the TCL block is higher than that in the two branches, which also validates that the inter-task learning can obtain better features. 

\begin{figure}[!htbp]
  \centering
  \includegraphics[width=\linewidth]{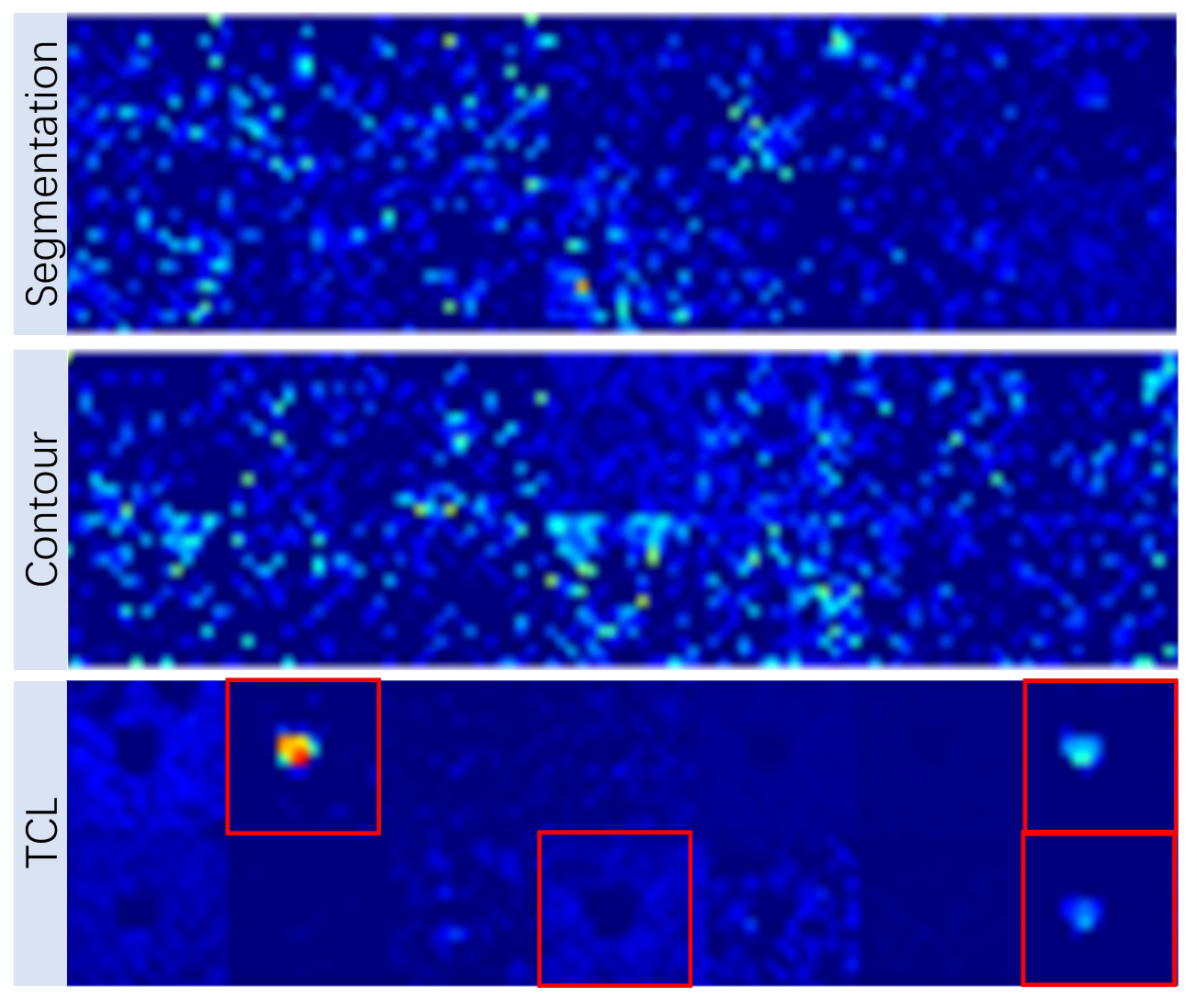}
  \caption{\label{fig:block5} Visualization of the features in the second top-level blocks of the segmentation branch (i.e., $branch_a$), the contour-aware branch (i.e., $branch_b$), and the TCL block. Due to limit of space, we convert the first 14 channels of the 3-D features (i.e., $64\times64\times64$) into 2-D map by channels for better visualization. The color from blue to red indicates low to high activity of the neurons.}
\end{figure}

Furthermore, Fig. \ref{fig:block5} show the features in the second top-level blocks. It can be seen that the features in the two branches cannot explicitly reveal the target organ. By combining information from the two branches, the highly activated neurons can still reveal a rough area of the target organ.


\subsubsection{The Effectiveness of Information Weight $\alpha$}
\begin{figure}[!htbp]
  \centering
  \includegraphics[width=\linewidth]{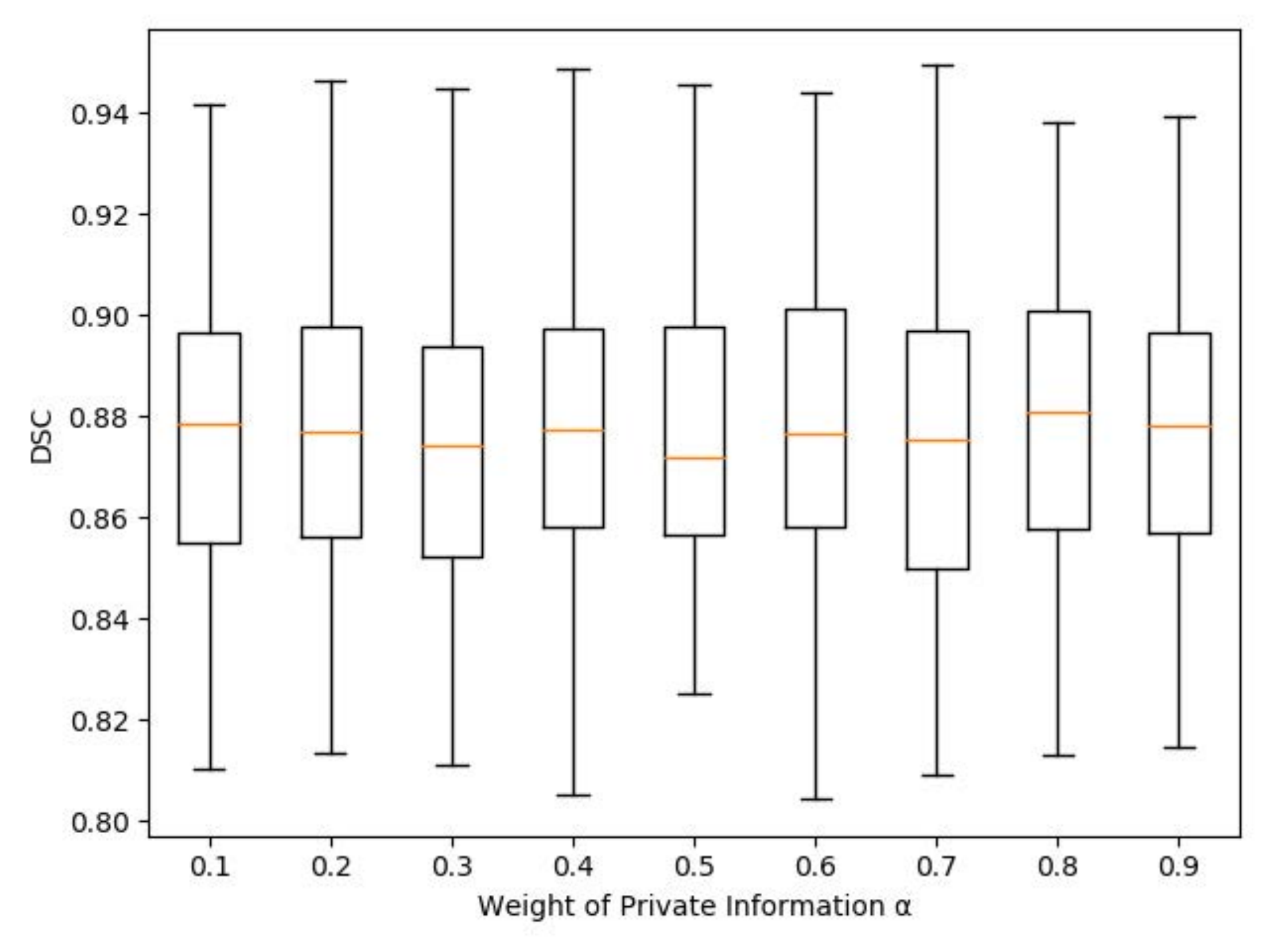}
  \caption{\label{fig:dscalpha}Box plot of the performance for HF-UNet-6 with respect to the weight alpha, using DSC.}
\end{figure}

\begin{figure}[!htbp]
  \centering
  \includegraphics[width=\linewidth]{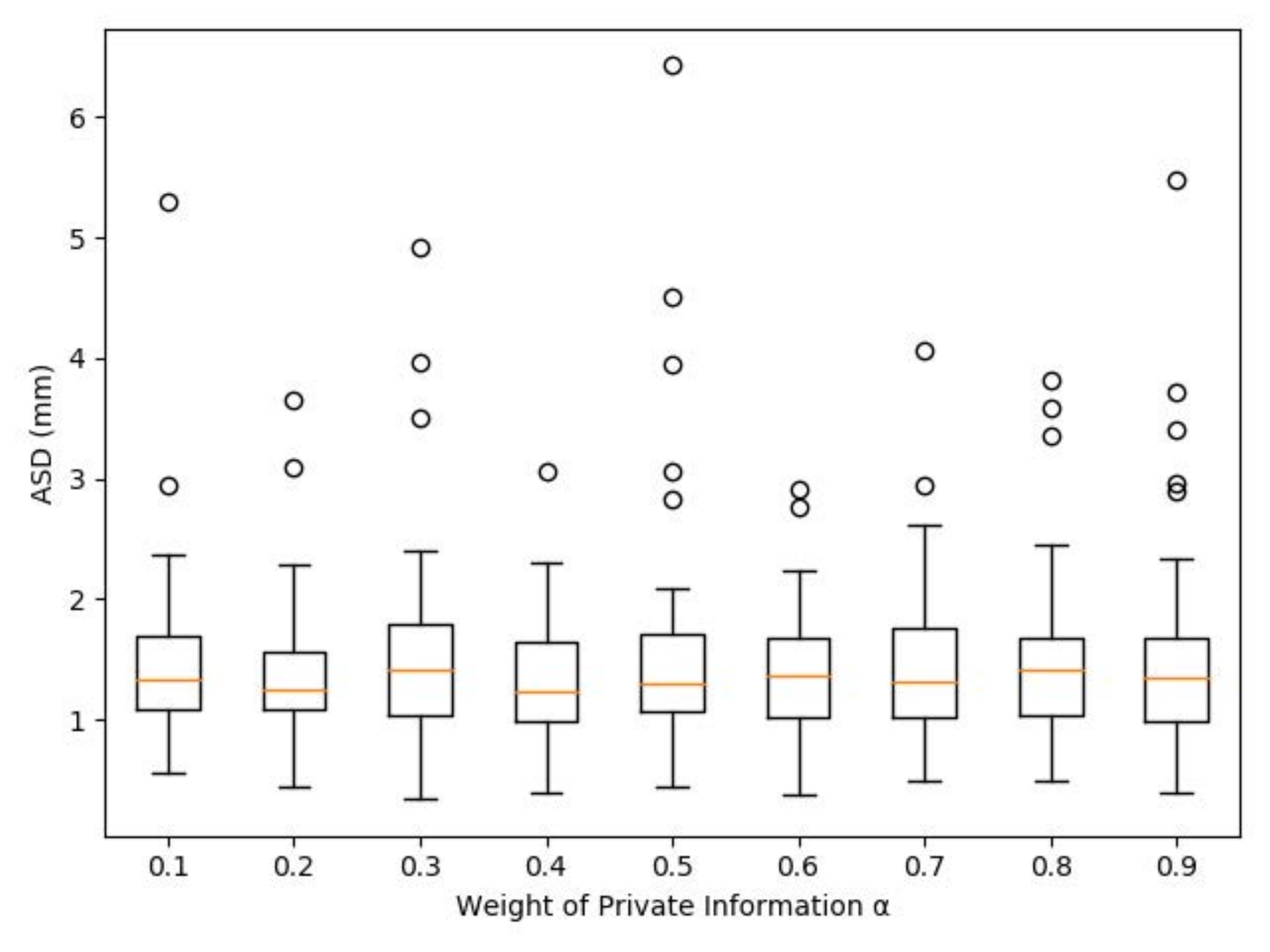}
  \caption{\label{fig:asdalpha}Box plot of the performance for HF-UNet-6 with respect to the weight alpha, using ASD.}
\end{figure}

The performance of HF-UNet with respect to different information weight $\alpha$ in DSC is shown in Fig. \ref{fig:dscalpha} in terms of ASD. \ref{fig:asdalpha}. For simplicity, we investigate discrete numbers of [0.1,0.9] for $\alpha$, with step 0.1. We do not  investigate the value $\{0,1\}$ of $\alpha$. HF-UNet will degrade to the late-branched network if $\alpha=0$ and will degrade to the dual-path learning network if $\alpha=1$. The figure suggests that the performance HF-UNet in terms of DSC is affected by $\alpha$. The HF-UNet with respect to $\alpha=0.8$ achieves the best performance in DSC of $0.880\pm0.028$. And the HF-UNet with respect to $\alpha=0.4$ achieves the best performance in ASD of $1.339\pm0.500$. However, when considering both DSC and ASD, the HF-UNet with respect to $\alpha=0.2$ achieves the best performance. Two conclusions can be made from these comparisons: (1) Balancing the information learned from the public and the private can achieve better performance; (2) The private information needs to be at a lower ratio (i.e., $\alpha=0.2$ in this experiment) to get better segmentation performance. Notably, designing an adaptive method for the ratio of $\alpha$ is important, and we leverage the attention mechanism in this work to achieve this goal.

\begin{figure*}[!htbp]
  \centering
  \includegraphics[width=0.9\linewidth]{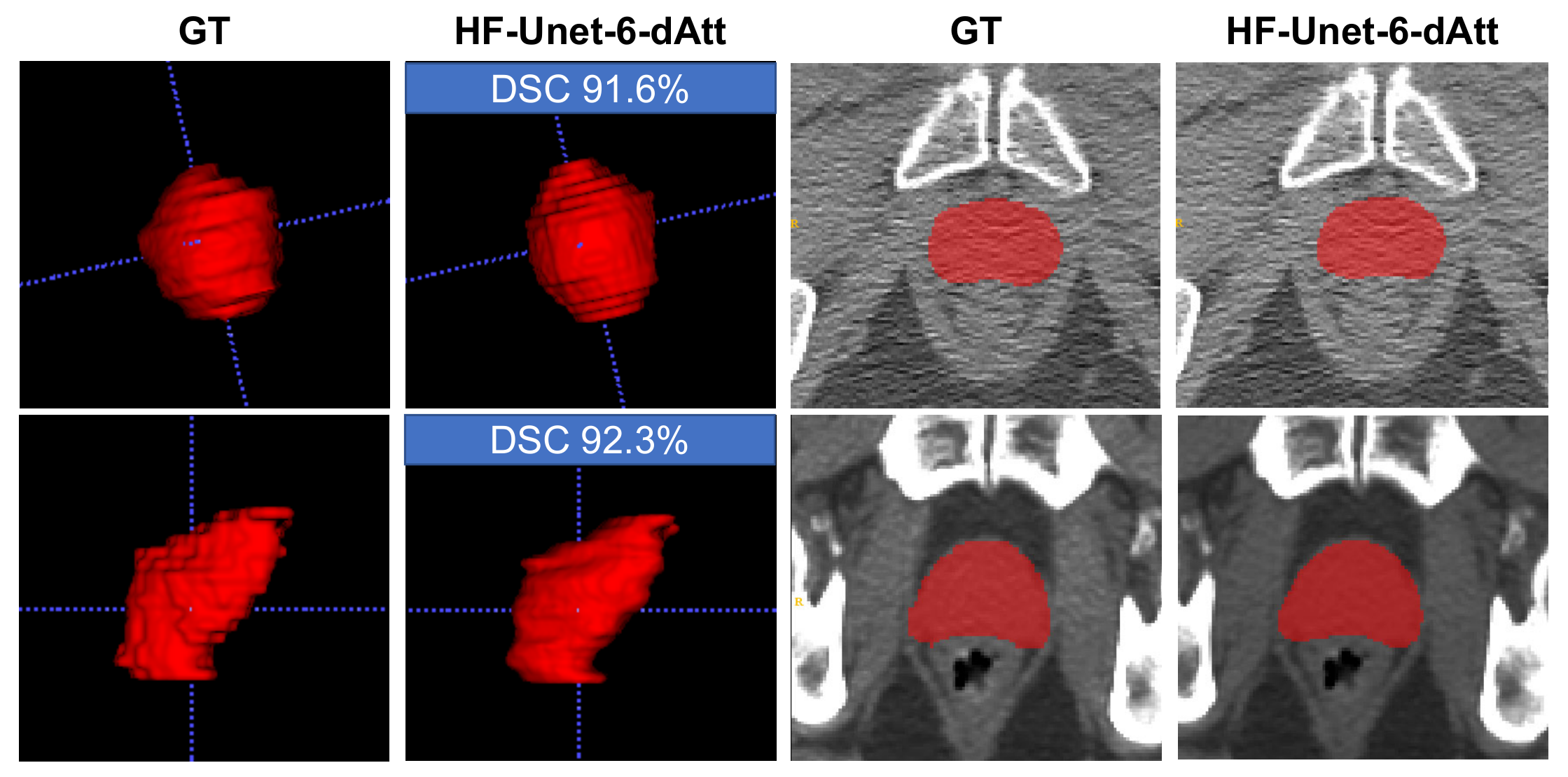}
  \caption{\label{fig:visseg} Visualization of typical cases of the prostate segmentation results of HF-UNet-6-dAtt. The smooth and high quality segmentation shown the ability of the proposed network.}
\end{figure*}

\begin{figure}[!htbp]
  \centering
  \includegraphics[width=\linewidth]{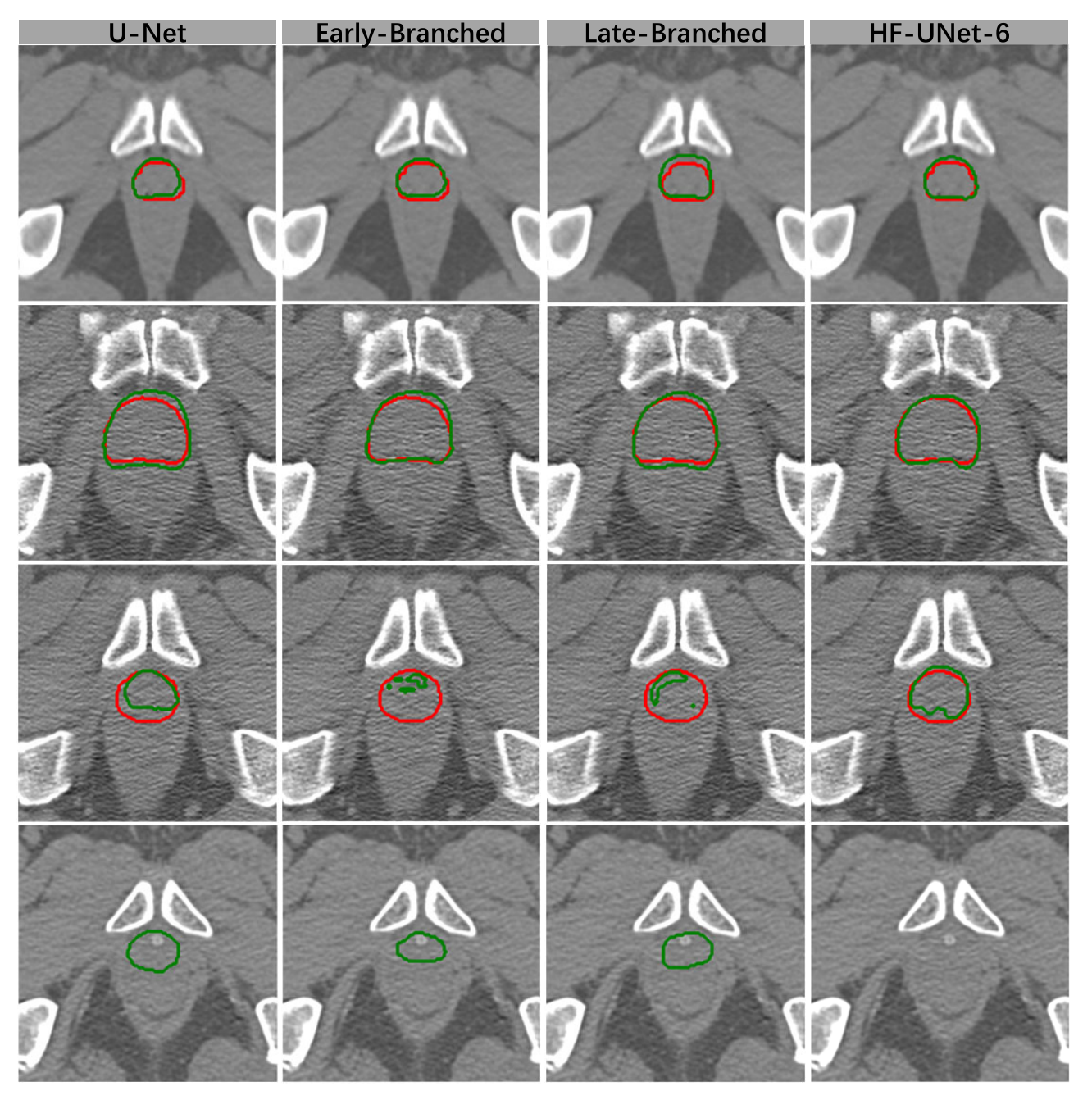}
  \caption{\label{fig:viscontour} Visualization of the prostate segmentation results. The method of U-Net, early-branched network, late-branched network and HF-UNet-6 are selected. Red contour indicates the groundtruth. Green Contour indicates the prediction.}
\end{figure}

\subsection{Comparison with State-of-the-art Methods}
We further compare our proposed HF-UNet-6 with several state-of-the-art segmentation methods \cite{martinez2014segmentation,shao2015locally,gao2015joint,gao2016accurate,ronneberger2015u,milletari2016v} that are mostly the deformable model-based and deep learning-based approaches. We briefly introduce these comparison methods as follows:
\begin{itemize}
    \item Martinez \etal \cite{martinez2014segmentation} proposed a Bayesian framework for deformable model-based segmentation.
    \item Shao \etal \cite{shao2015locally} proposed to regress the boundary of the prostate for more accurate segmentation of the prostate.
    \item Gao \etal \cite{gao2015joint,gao2016accurate} proposed deformable model-based segmentation methods for pelvic organs, which combine the guidance of the displacement maps of the organs.
    \item Ronneberger \etal \cite{ronneberger2015u} proposed a popular U-Net with coordinate feature extraction and the expansion paths with skip-connections.
    \item Milletari \etal \cite{milletari2016v} proposed the V-Net, which is a U-Net like network structure with residual connections in the convolutional blocks.
    \item He \etal \cite{he2019pelvic} proposed a two-stage U-Net based network to leverage boundary information to improve segmentation performance.
\end{itemize}

It is worth noting that several competing methods also leverage the boundary information (\eg, regression cues \cite{shao2015locally} or displacement maps \cite{gao2015joint,gao2016accurate}) for performing segmentation, which obtained superior performance in the past. Their experimental results are given in Table \ref{Table:EvaPro}. 
To evaluate the performance of the proposed attention-based TCL block, we build three HF-UNet-6 networks with the aforementioned attention mechanism, \ie, 1) HF-UNet-6-pAtt with the position-wise attention, 2) HF-UNet-6-cAtt with the channel-wise attention; and 3) HF-UNet-6-dAtt with both the position-wise and channel-wise attention (see in Fig. \ref{fig:aTCLblock}).
One can observe from Table \ref{Table:EvaPro} that HF-UNet-6-dAtt consistently achieves the best performance in terms of both DSC and ASD, which indicating the efficacy of the proposed method. Also, we would like to mention that, along with the improvement relatively saturated on DSC, our method could improve ASD with a large margin (over $60\%$) compared with U-Net, from $3.89$mm to $1.31$mm, and more than $30\%$ over the state-of-the-art method, from $1.71$mm to $1.31$mm. This indicates that the proposed method can generate a more robust shape of the prostate mask, which is closer to the actual organ surface. This is mainly due to the incorporation of morphological representation, which helps better delineate the organ boundary. Also, the proposed hierarchically fusion strategy can bring detailed contour information from the low- to high-level features of the network.

\subsection{Visualization Results}

We show the visualization results of segmentation produced by our proposed HF-UNet-6 and three competitors (i.e., UNet, early- and late-branched networks) in Fig. \ref{fig:viscontour}. It can be observed that HF-UNet-6 shows the best performance among the four comparison methods.

The visualization comparison of the contours of the four methods is shown in Fig. \ref{fig:viscontour}. Note that HF-UNet-6 can implicitly leverage contour information, and thus can better delineate prostate contours in most cases of CT images. Therefore, we only visualize results of some very challenging cases in Fig. \ref{fig:viscontour}, from which we can observe that our HF-UNet-6 method consistently outperforms other networks for these challenging cases. 
The first two rows show that HF-UNet-6 generates higher overlapping contours with the groundtruth, and other networks generate less-accurate prostate contours in several cases. 
In some very challenging cases, the three competing methods generate under-segmentations in the third row and while over-segmentations in the last row. By comparison, the proposed network performs consistently better in these cases.
These results imply that the proposed HF-UNet can work well in the challenging cases with very low tissue contrast.
The visualization for the segmentation results of the proposed HF-UNet-6-dAtt on two typical cases is provided in Fig. \ref{fig:visseg}, indicating the robustness of segmentation across subjects by the proposed method.

\section{Discussion and Conclusion}

In this paper, we propose a hierarchically fused multi-task fully convolutional network (i.e., HF-UNet) for segmentation of prostate in CT images, to learn better task-specific features. Specifically, the informative contour sensitive label is proposed to enhance the capacity of the network to better identify prostate boundaries.
HF-UNet is proved to preserve task-specific features in multi-task learning.
And the proposed Task Consistency Learning (TCL) block allows cross-task feature sharing at each level.
Moreover, we investigate different configurations of the network with a different number of TCL blocks.
Moreover, instead of the whole-shared strategy in the previous networks, we explore the ratio of shared information and task-specific information, which can promote the performance of multi-task networks. 
The ratio of the two kinds of information used in HF-UNet can change the performance of the network. Therefore, we proposed an attention-based residual mechanism with both position-wise attention and channel-wise attention to learn the optimal ratio of the information, which improves segmentation performance by the proposed method.
Our evaluations on the 339-patient CT image dataset showed the effectiveness of the proposed HF-UNet, which outperforms the state-of-the-art methods. 
Moreover, our framework can be easily applied to other contour sensitive segmentation tasks.
The proposed network is generally enough to be applied to other applications with multiple tasks. Besides, we can also design different architectures for different branches to deal with different tasks in different application cases, which will be our future work.

\bibliographystyle{IEEEtran}
\bibliography{IEEEabrv,hffcn}




\ifCLASSOPTIONcaptionsoff
  \newpage
\fi

\end{document}